\def\U{\Upsilon}

\def\z{\zeta}
\def\u{\underline}
\def\o{\overline}
\def\T{\Theta}
\def\a{\alpha}
\def\b{\beta}
\def\g{\gamma}

\def\d{\delta}
\def\3{  {1 \over {3!}}  }
\def\uh#1{\rlap{\lower1ex\hbox{$*$}}#1{}}
\def\ut#1{\rlap{\lower1ex\hbox{$\sim$}}#1{}}
\def\uut#1{\rlap{\lower2ex\hbox{$\sim$}}#1{}}
\def\oh#1{\rlap{\raise1ex\hbox{$*$}}#1{}}
\def\ot#1{\rlap{\raise1ex\hbox{$\sim$}}#1{}}
\def\ooh#1{\rlap{\raise2ex\hbox{$\thinspace *$}}#1{}}
\def\oot#1{\rlap{\raise1.5ex\hbox{$\sim$}}#1{}}
\def\RR{\hbox{$I$\kern-3.8pt $R$}}
\def\sss{\scriptscriptstyle}
\def\be{\begin{equation}}
\def\ee{\end{equation}}
\documentstyle[prd,preprint,eqsecnum,aps]{revtex}
\begin{document}
\tighten
\preprint{\vbox{\baselineskip=12pt
\rightline{UCSBTH-95-26}
\rightline{CTMP/012/NCSU}
\rightline{gr-qc/9509026}}}
\title{On Relativistic Material Reference Systems}
\author{J. David Brown\cite{david}}
\address{Department of Physics and Department of Mathematics\\
North Carolina State University, Raleigh, NC 27695--8202}
\author{Donald Marolf\cite{don}}
\address{Department of Physics\\
The University of California, Santa Barbara, CA 93106--9530}
\date{September, 1995}
\maketitle
\begin{abstract}
This work closes certain gaps in the literature on material
reference systems in general relativity.  It is shown
that perfect fluids are a special case of DeWitt's
relativistic elastic media and that the velocity--potential
formalism for perfect fluids can be interpreted as describing a
perfect fluid coupled to a fleet of clocks.
A Hamiltonian analysis of the elastic media with clocks
is carried out and  the constraints that
arise when the system is coupled to gravity are studied.  When
the Hamiltonian constraint is resolved with respect to the
clock momentum, the resulting true Hamiltonian is found to be
a functional only of the gravitational variables. The true
Hamiltonian is explicitly displayed when the medium is dust,
and is shown to depend on the detailed construction of the
clocks.
\end{abstract}
\pacs{???}
\section{Introduction}

The use of material reference systems in general
relativity has a long and noble history.  Beginning with the systems
of rods and clocks conceived by Einstein \cite{Ein} and
Hilbert \cite{Hil}, material systems have been used as a
physical means of specifying events in
spacetime and for addressing conceptual questions in classical gravity.
That such systems also provide important tools for
{\it quantum} gravity was pointed out by DeWitt \cite{Bryce},
who
used them to analyze the implications of the uncertainty principle
for measurements of the gravitational field.

The original systems of rigid rods and massless clocks discussed
by Einstein and Hilbert represent unphysical idealizations. Since their
time,
attempts have been made
to remedy this shortcoming by developing a more physically
realistic description of the reference medium. While still a
phenomenological description, a dynamical reference system consistent
with relativity may be found in the clocks and elastic media studied
by DeWitt \cite{Bryce}.  In the same spirit, perfect
fluids \cite{pf1} have been employed as reference
systems in the quantization of various model problems in gravity
\cite{LDM}.
More recently, Kucha\v{r} {\it et al.\/} \cite{KST} have developed a
scheme for incorporating reference systems in general relativity
through the introduction of coordinate conditions \cite{IK}. For the
cases examined thus far, the reference materials that arise through
this approach have certain unphysical properties. This
has motivated the investigation of pressureless perfect
fluid (dust), which is unrelated to any obvious coordinate condition,
as a phenomenological but physically realistic reference system \cite{BK}.

Here we pursue the modest goal of closing certain gaps in
the literature on
reference materials.  First, we establish in Sec.~II the connection
between the elastic media of DeWitt \cite{Bryce} and the relativistic
perfect fluids of Refs.~\cite{pf1,pf2,d1}, which include dust as a
special case. In particular we show that perfect
fluids are equivalent to elastic media when the later are
homogeneous and
isotropic.

We
then turn to the study of reference {\it systems} defined by the
coupling of clocks to
reference {\it materials}. In
Sec.~III.A we
demonstrate that a perfect fluid, as described by the action $S$ of
Refs.~\cite{d1,d2} for the isentropic case, can be interpreted as a
perfect fluid {\it coupled to a fleet of clocks} (with the details
presented in Appendix A). We show the inequivalence of this
clock coupling to the coupling used by DeWitt \cite{Bryce}, and
discuss the advantages of DeWitt's method. We also point out that
for a nonisentropic perfect fluid $S$, the thermasy \cite{Dantzig}
can be reinterpreted as a clock variable with DeWitt--type
coupling. Then, in preparation for Sec.~IV,  we
perform  in Sec.~III.B a Hamiltonian analysis of the reference system
formed by adding clocks (in the manner of DeWitt) to
the elastic medium.  This proceeds along the lines of that done in
Refs.~\cite{d1,d2} for perfect fluids.

In Sec.~IV we study the coupling of the reference system to gravity
and the resulting canonical constraints. When the Hamiltonian constraint
is resolved with respect to the clock momenta, we find that
the true Hamiltonian depends only on the gravitational variables, not on
the clock or particle variables. (Some details are derived in Appendix
B.) The case of dust is particularly simple, and we explore the
use of more general clocks than those used in Ref.~\cite{BK}.
For this medium and for
appropriately designed clocks, the constraints
can be resolved with respect to the clock momenta using only
analytic operations (i.e., without taking a square
root). This allows us to bypass a technical difficulty when defining
the quantum theory \cite{BK} although, as we point out,
interpretational difficulties appear in its place.

We  use the following notation, which is consistent with
that of Refs.~\cite{d1,d2}.
The action for a reference {\it system}, which includes both matter
and clocks, will be denoted by ${S}$, while the action for
the corresponding reference material alone, without clocks,  will
be denoted by $\bar S$.  Thus, the addition of clocks
changes a `barred' system into an `un--barred' system.

\section{Elastic Media and Perfect Fluids}

In this section we address the relation
between DeWitt's elastic media and
the perfect fluids of Refs.~\cite{pf2,d1,d2}. In fact, the main
difference is that the elastic medium is presented in the
Lagrangian picture while the perfect fluids are typically
presented in the Eulerian picture. Thus, we first review
the elastic medium in the original Lagrangian description, and
then rewrite it in the Eulerian picture.  The perfect fluid
action is then recognized to be a special case of the action
for an elastic medium.

We use
the terms Lagrangian and Eulerian in the following way.
The term ``Lagrangian picture" refers to the description
in which the basic
variables tell the spatial location of a given particle, or the
spacetime location of a given event on a given particle world line.
The term ``Eulerian picture" refers to the description in which the basic
variables tell which particle resides at a given spatial location,
or which particle passes through a given spacetime event.

\subsection{A Review of Elastic Media}
We now begin with a summary of DeWitt's elastic media \cite{Bryce}.
A single free relativistic particle moving in a spacetime
${\cal M} = \Sigma \times\RR$ with metric $\gamma_{\alpha \beta}$
can be described by the action
\begin{equation}
{\bar S}_1[\Upsilon^\alpha;\gamma_{\a\b}] = -\int d\sigma \,
m\sqrt{ - \dot{\Upsilon}^{\alpha}
\dot{\Upsilon}^{\beta}\gamma_{\alpha\beta}(\Upsilon)} \ .
\end{equation}
Here, $\sigma$ is an arbitrary parameter along the particle world
line and $y^\alpha = \Upsilon^\alpha(\sigma)$ are the spacetime
coordinates of the particle.
Also, $m>0$ is the mass of the particle and the
dot denotes a derivative with respect to $\sigma$.  The semicolon
notation in $\bar{S}_1[\Upsilon^\alpha;\gamma_{\alpha \beta}]$ indicates
that this action is to be varied with respect to $\Upsilon^\alpha$,
with $\gamma_{\alpha \beta}$ treated as a background field.

As in Ref.~\cite{Bryce}, we may also consider fleets of such particles
and we may add local interactions.
If the particles are labeled by a set of Lagrangian coordinates
$\zeta^i$, $i\in\{1,2,3\}$, then such a system can be described by the
action
\begin{equation}
\label{EMLL}
\label{fleet}
\bar{S}_{\sss LL}[\Upsilon^\alpha;\gamma_{\a\b}] =
\int d\sigma \int_{\cal S} d^3\z
\biggl\{ - (\u{n}m + \u{w}) \sqrt{ -
\dot{\Upsilon}^{\alpha}
\dot{\Upsilon}^{\beta}\gamma_{\alpha\beta}(\Upsilon) }\biggr\} \ ,
\end{equation}
where ${\cal S}$ is the ``matter space" manifold \cite{pf2,d1}
whose points $\zeta\in{\cal S}$ label the particle worldlines.
In Eq.~(2.2), the dynamical variables $\Upsilon^\alpha$
are functions of $\sigma$ and $\zeta^i$. The
quantity $\u{n}$ is the particle number density, so that
$\u{n}\, d^3\zeta$ is the number of particles in the coordinate
cell $d^3\zeta$. The quantity $\u{w}$ is the interaction energy
density in the co--moving frame, so that
$\u{w}\, d^3\zeta$ is the interaction energy in the coordinate cell
$d^3\zeta$ as measured in the rest frame of the particles.
In order to make the transformation properties of these
expressions clear, we explicitly indicate density weights with
respect to changes of the coordinates $\zeta^i$ on ${\cal S}$ with
underlines. Thus, $\u{n}$ and $\u{w}$ are densities in
the matter space ${\cal S}$.

The functions $m$, $\u{n}$, and $\u{w}$ can depend explicitly
on $\zeta^i$ and, in addition, the interaction energy
$\u{w}$ can have an ultralocal dependence on the matter space metric
(``fleet metric'') $h_{ij}$. The fleet metric is
defined such that $ds =\sqrt{ h_{ij}d\zeta^i d\zeta^j}$ measures the
orthogonal distance
$ds$ between neighboring world lines with
Lagrangian coordinates $\zeta^i$ and $\zeta^i + d\zeta^i$.
By including an explicit dependence on $\zeta^i$ in the mass
$m$ and interaction energy $\u{w}$, we allow for the possibility that
the particles are not identical.
In terms of the matter four--velocity
\begin{equation}
U^\a = { { \dot{\U}^\a} \over {\sqrt { - \dot{\U}^\b
\dot{\U}^\d\gamma_{\b\d}  }} } \ ,
\end{equation}
the matter space metric takes the form
\begin{equation}
h_{ij} = \U^{\alpha},_i (\gamma_{\alpha \beta}
+ U_{\alpha}U_{\beta}) \U^{\beta},_j
\end{equation}
where the commas denote derivatives with respect to $\z^i$.
In general $h_{ij}$ will not be the metric of any hypersurface of the
spacetime manifold ${\cal M}$.

The system described by the action $\bar{S}_{\sss LL}$ of
Eq.~(\ref{fleet}) is referred to as
an {\it elastic medium}. The subscript $LL$ indicates that it is the
Lagrangian form (as opposed to the Hamiltonian form) of the action in
the Lagrangian picture. Observe that this system is
reparametrization invariant. That is, $\bar{S}_{\sss LL}$ is invariant under
the transformation $\delta\Upsilon^\alpha = -{\dot\Upsilon}^\alpha
\epsilon$ induced by a reparametrization
$\sigma\to\sigma + \epsilon(\sigma,\zeta)$ of the particle
worldlines, where $\epsilon$ vanishes at the endpoints in $\sigma$.

\subsection{The Connection to Fluids}

We will now derive the Eulerian description of the elastic medium, in
which the action is written as an integral over arbitrary spacetime
coordinates $y^{\alpha}$ on ${\cal M}$, and show its relations to
the perfect fluids of
Refs.~\cite{pf2,d1,d2}.

To begin, let us assume that
in the region of spacetime described, one and only one particle of the
medium passes through each event.
Then the Lagrangian coordinates $\zeta^i$ along with the
worldline parameters $\sigma$ form a set of coordinates in
the spacetime region. These coordinates are related to the
coordinates $y^\alpha$ by the mappings $y^\alpha =
\Upsilon^\alpha(\sigma,\zeta)$. We can introduce the inverse
mappings
\begin{equation}
\sigma = Z^0(y) \ ,\qquad \zeta^i = Z^i(y) \ ,
\end{equation}
such that $y^\alpha = \Upsilon^\alpha(Z^0(y),Z^i(y))$ is
an identity. Note that $Z^i(y)$ give the labels $\zeta^i$ of the
particle present at the event $y^{\alpha}$. The transition from the
Lagrangian picture to the Eulerian picture is obtained by a change
of dynamical variables in
which $\Upsilon^\alpha(\sigma,\zeta)$ is replaced by
$Z^0(y)$ and $Z^i(y)$.

To perform this change of variables, we first calculate the Jacobian
$|{\partial(\sigma,\zeta)} / {\partial y}|$.
Let us assume that the coordinate system $\sigma$, $\zeta^i$ has the
orientation of ${\cal M}$. Then
we have the identity
\begin{eqnarray} \label{wedge}
d\sigma \wedge d\z^1 \wedge d\z^2 \wedge d\z^3 & = &
Z^0,_{\alpha} Z^1,_{\b} Z^2,_{\g} Z^3,_{\d} dy^{\a} \wedge
dy^{\b} \wedge dy^{\g} \wedge dy^{\d} \nonumber\\
& = & -\frac{1}{3!}\frac{\sqrt{-\gamma}}{\sqrt{h}} \epsilon^{\a\b\g\d}
\epsilon_{ijk} Z^0,_{\alpha} Z^i,_{\b} Z^j,_{\g} Z^k,_{\d} dy^{0} \wedge
dy^{1} \wedge dy^{2} \wedge dy^{3}  \ ,
\end{eqnarray}
where the commas followed by greek letters denote derivatives with
respect to $y^\a$. Here, $\epsilon^{\a\b\g\d}$ is
the totally antisymmetric contravariant tensor on ${\cal M}$ with
$\epsilon^{0123} = -1/\sqrt{-\gamma}$, and $\gamma$ is
the determinant of the spacetime metric $\gamma_{\a\b}$. Similarly,
$\epsilon_{ijk}$ is the totally antisymmetric covariant tensor on
${\cal S}$ with $\epsilon_{123} = \sqrt{h}$.
Observe that the inverse fleet metric can be written as
\begin{equation}
\label{ifm}
h^{ij} =  Z^i,_{\alpha}
\gamma^{\a \b} Z^j,_{\b} \ ,
\end{equation}
so that $h$ can be expressed in terms of the Eulerian
variables as $h = 1/\det(h^{ij})$.
It will also be convenient to express the particle four--velocity $U^\a$
in terms of the new variables:
\begin{equation}
\label{4v}
U^\a = -(1/3!) \epsilon^{\a\b\g\d} Z^i,_\b Z^j,_\g Z^k,_\d
\epsilon_{ijk} \ .
\end{equation}
One can see that this expression
is indeed the particle velocity by verifying
that $U^\a Z^i,_\a = 0$ and $U^\a U_\a = -1$.
This allows us to rewrite the measure (\ref{wedge}) as
\begin{equation}
d\sigma \wedge d\z^1 \wedge d\z^2 \wedge d\z^3
= \sqrt{-\gamma/h}\, Z^0,_{\alpha} U^\a dy^{0} \wedge
dy^{1} \wedge dy^{2} \wedge dy^{3} \ .
\end{equation}

Finally, note
that ${\dot\Upsilon}^\a$ is proportional to $U^\a$ and
${\dot\Upsilon}^\a Z^0,_\a = 1$, so that ${\dot\Upsilon}^\a
= U^\a/(U^\b Z^0,_\b)$.
Combining this with
the above results, we find that the
action (2.2)
takes the form
\begin{equation}
\label{EMLE}
\bar{S}_{\sss LE}[Z^i;\gamma_{\a\b}]  = - \int_{\cal M} d^4y
 \sqrt{-\gamma/h}
(\u{n} m + \u{w})  \ .
\end{equation}
in the Eulerian picture.
Here, the mass $m$ and number density $\u{n}$ are fixed functions of
$Z^i$ while the interaction energy $\u{w}$ is a fixed function of
$h_{ij}$ and $Z^k$.  The fleet metric $h_{ij}$ is taken to be a function
of the spacetime metric $\gamma_{\a\b}$
and the variables $Z^i$ through Eq.~(\ref{ifm}).

Observe that the Eulerian form (\ref{EMLE}) of the action does not
depend on the variable $Z^0$. This is a consequence of reparametrization
invariance: A reparametrization of the world lines induces
the transformation $\delta Z^0 = \epsilon$ while leaving the other
variables $Z^i$ alone. The action (2.10) is invariant
precisely because it is independent of $Z^0$.\footnote{There is a
subtle point here. The action (2.2) is defined for a fixed
integration region in ${\cal S}\times\RR$; that is, for fixed ranges
of the integration parameters $\sigma$ and $\zeta^i$. The integration
region in ${\cal M}$ is determined by the integration region in
${\cal S}\times\RR$ only if we fix $\Upsilon^\a$ at the boundaries.
Then, in particular, the endpoints in $\sigma$ determine initial and
final hypersurfaces in ${\cal M}$ which we assume to be spacelike.
Since the range of $\sigma$ in
(2.2) is fixed, the action functional (2.10) is defined for the class
of variables $Z^0(y)$ with fixed values on the initial and final
hypersurfaces in ${\cal M}$. Therefore the gauge freedom in (2.10)
consists of variations $\delta Z^0 = \epsilon$ for which $\epsilon$
vanishes on the initial and
final hypersurfaces. In this way we see that the gauge freedom for the
actions (2.2) and (2.10) coincide.} This gauge freedom is
removed simply by dropping $Z^0$ from the list of dynamical
variables. Thus, we view the action (2.10) as a functional
of $Z^i$ (and $\gamma_{\alpha\beta}$) only.

{}From the action (2.10), it is straightforward to show
that the isentropic perfect fluid
action $\bar{S}$ given by Eq.~(6.15) of Ref.~\cite{d1}
(or the isentropic
case of Eq.~(4.20) of Ref.~\cite{d2}) is equivalent to
a ``homogeneous and isotropic elastic medium."
To do so,
consider the case in which the mass
$m$ is independent of the particle labels $Z^i$, so the
particles are identical. Also assume that the {\it proper\/}
interaction energy density $w = \u{w}/\sqrt{h}$ (which is the
interaction energy per unit proper spatial volume as measured
in the rest frame of the matter) depends only on the {\it proper\/}
particle number density $n=\u{n}/\sqrt{h}$ (which is the number of
particles per unit proper spatial volume as measured in the
rest frame of the matter). That is, $\u{w}$ depends
on $Z^i$ and $h_{ij}$ only through the combination
$\u{w} = \sqrt{h} w(\u{n}/\sqrt{h})$ for some function $w(n)$.

The factor $(\u{n}m + \u{w})/\sqrt{h}$ that appears in the
integrand of the action $\bar{S}_{\sss LE}$ is the {\it proper\/}
energy density of the medium, which we will denote by $\rho$. Our
restriction to a homogeneous and isotropic medium implies that $\rho$
only depends on  the proper number density, $\rho = \rho(n)$. Thus,
the action (\ref{EMLE}) in the homogeneous, isotropic case may be
written as
\be
\label{fluid}
\bar{S}_{\sss LE}[Z^i;\gamma_{\alpha\beta}] =
-\int_{\cal M} d^4y \sqrt{-\gamma} \rho(n) \ ,
\ee
where $n=\u{n}/\sqrt{h}$. This is precisely the perfect fluid action
$\bar S$ given in Refs.~\cite{d1,d2} for the isentropic case, although
in those references the number density $n$ was expressed somewhat
differently. To see that the definition of $n$ in Refs.~\cite{d1,d2}
agrees with the present definition, we use Eq.~(2.7) to
write the number density explicitly as
\be
\label{nofZ}
n = \u{n}/\sqrt{h} = \biggl[ {1 \over {3!}} \u{n}^2 \o{\epsilon}_{ijk}
\o{\epsilon}_{i'j'k'} Z^i,_\a Z^{i'},_{\a'} \gamma^{\a \a'}
Z^j,_\b Z^{j'},_{\b'} \gamma^{\b \b'} Z^k,_\delta Z^{k'},_{\delta'}
\gamma^{\delta \delta'}\biggr]^{1/2} \ .
\ee
Here, $\o{\epsilon}_{ijk}$ is the antisymmetric tensor density
(of weight $-1$) on ${\cal S}$ with $\o{\epsilon}_{123} = 1$. If we
identify the tensor
$\eta_{ijk}(\zeta)$ of Refs.~\cite{d1,d2} with
$\u{n}(\zeta) \o{\epsilon}_{ijk}$, then the expression (\ref{nofZ})
exactly matches the definition of $n$ given through Eq.~(5.1) of
\cite{d1} (and through Eqs.~(4.23) and (4.26) of \cite{d2}).
As a result, the (`barred') one component
isentropic perfect fluid of Refs.~\cite{d1,d2}
is seen to be a special case of the
relativistic elastic medium of DeWitt \cite{Bryce}.

\section{Clocks and Reference Systems}

Reference materials such as elastic
media and perfect fluids can be used
to provide a physical system of coordinates in space. However, by itself,
a reference material does not provide a complete coordinate system
in space{\it time},
as all points along a given particle worldline are
labeled by the same coordinates $\zeta^i$. This can be remedied by
adding an additional degree of freedom to the particles
whose value changes along the worldlines.
Such a degree of freedom may be called a `fleet of clocks' and
a reference medium coupled to a fleet of clocks is
said to constitute a {\it reference system}.  Note that,
so far, we have not distinguished between `good clocks'  which
accurately measure proper time and `bad clocks' whose readings
vary along the worldlines in a more complicated way.

\subsection{Coupling clocks}

The literature contains two different mechanisms for
coupling additional `clock' degrees of freedom to a reference material.
One of these was used by DeWitt \cite{Bryce} and was explicitly
described as a coupling of clocks to an elastic medium.
The other is implicitly contained in the literature
on isentropic (single component) perfect fluids \cite{pf1,d1,d2},
although the word `clock' does not appear in any of these works.
Moreover, with a reinterpretation of variables and a suitable choice of
equation of state, a {\it non}isentropic perfect fluid is actually
equivalent to a homogeneous, isotropic elastic medium with the
DeWitt--type clock coupling.
The following summary along with the results in Appendix \ref{+clocks}
should clarify this situation.

We first restrict our attention to isentropic perfect fluids.
Historically, two different action principles were developed for
relativistic perfect fluids which both used scalar fields (``velocity
potentials") as the basic variables. These actions were later shown to be
equivalent \cite{d2}. In particular, the action of Ref.~\cite{pf1},
which we will denote by $S$,
was shown to differ from the action of Ref.~\cite{pf2}, which
is denoted by $\bar S$ in Eq.~(2.11),
by the addition of one degree of freedom per space point. This
additional degree of freedom is cyclic, so the action $\bar S$ can
be derived from $S$ by removing the extra degree of freedom through
Routh's procedure \cite{Goldstein}.

Since the `un--barred' description of perfect fluids contains an
extra degree of freedom per space point, and since this degree
of freedom changes along the worldlines,
the philosophy stated at the beginning of this section allows us to
interpret this degree of freedom as representing a fleet of clocks.
For this reason, we refer to the `un--barred' fluid as a fluid
{\it coupled to a fleet of clocks}.

For both the clock coupling discussed by DeWitt and the one
implicit in the perfect fluid literature, the
essential idea is to add a pair $(\Theta, J)$ of first order degrees of
freedom for each particle in the medium.
Thus, we consider two fields, $\Theta(\sigma,\zeta)$
and $J(\sigma,\zeta)$, and add the first order kinetic term
\be
\label{kinetic}
\int d\sigma \int_{\cal S} d^3\zeta \, {\underline n}
J \dot\Theta
\ee
to the Lagrangian picture action (\ref{EMLL}).
The Eulerian form of the kinetic term may be
obtained from Eq.~(2.9) and the
relation $\delta^\alpha_\beta = {\dot\Upsilon}^\alpha Z^0,_\beta
+ \Upsilon^\alpha,_i Z^i,_\beta$ and is given by
\begin{equation}
\label{Ekin}
\int_{\cal M} d^4y \,
\u{n} J \Theta,_{\alpha} U^\alpha \sqrt{-\gamma/h} \ .
\end{equation}
The clocks are then
coupled to the particles by letting  either $\u{n}(\zeta)$ or $m(\zeta)$
depend on $J$ in the original action.

To produce the `un-barred' fluid action of Ref.~\cite{pf1,d1,d2},
the clocks are coupled by letting $\u{n}(\zeta) \to J\u{n}(\zeta)$
in the action (2.11) and adding the kinetic term (3.2).
A detailed explanation of how this produces the
`un-barred' action of Ref.~\cite{pf1,d1,d2} is
given in Appendix \ref{+clocks}.
This method of clock coupling does not, in general, lead to a
`good' set of clocks. This can be seen from the equation of motion
for $J$, which is
\begin{equation}
\label{fluidclocks}
\Theta,_\alpha U^\alpha = \rho'(J\u{n}/\sqrt{h})
\end{equation}
where $\rho'(n) = \partial\rho(n)/\partial n$. The relation
(\ref{fluidclocks}) shows that
the rate of advance of the clock variable $\Theta$ relative to the
flow of proper time may depend on the particle density $n$.
Thus, for clocks coupled in this way,
the internal clock mechanism is not shielded from the external pressures
and forces between particles. For some equations of state $\rho(n)$,
such clocks may
not even run monotonically in proper time.

Because the method $\u{n} \to J\u{n}$ of clock coupling does not in
general yield a good set of clocks, we will focus on the method
described by DeWitt \cite{Bryce}.
In this method the mass $m$ is allowed to depend on
$J$.  Thus, the
mass of each particle is no longer a fixed constant and, in fact, it
acts as a Hamiltonian $m(\zeta^i,J)$ that drives the motion of the
clock $\Theta(\zeta^i)$ attached to particle $\zeta^i$.  In the
Lagrangian picture, the action
$S_{\sss LL}[\Upsilon^\alpha,J,\Theta;\gamma_{\alpha\beta}]$ is obtained
by adding
Eqs.~(2.2) and (3.1) and
taking
the number density,
mass, and internal energy density
to depend on $\zeta^i$, $J$, and $h_{jk}$ through
$\u{n} = \u{n}(\zeta^i)$,
$m = m(\zeta^i,J)$, and $\u{w} = \u{w}(\zeta^i,h_{jk})$.
The action also takes a simple form in the Eulerian picture:
\begin{equation}
\label{SLE}
S_{\sss LE}[Z^i,J,\Theta;\gamma_{\alpha\beta}] = \int_{\cal M} d^4y
\sqrt{-\gamma/h} \Bigl\{ \u{n}J\Theta,_\alpha U^\alpha -
(\u{n} m+ \u{w}) \Bigr\} \ ,
\end{equation}
where $\u{n} = \u{n}(Z^i)$, $m = m(Z^i,J)$, and $\u{w} =
\u{w}(Z^i,h_{jk})$.
Note that the
equation of motion for $J$ shows that the clock
satisfies
\begin{equation}
\label{Bclocks}
\Theta,_\alpha U^\alpha = \partial m/\partial J \ ,
\end{equation}
while
the equation of motion for $\Theta$ shows that $J$
(and therefore
${{\partial m} / {\partial J}}$)
is a constant
along each particle world line.
Thus, for the DeWitt--type coupling, $\Theta$ increases in direct
proportion to proper time along the worldline and always defines a good
clock. We see that
(\ref{fluidclocks}) and (\ref{Bclocks})
coincide
when the medium is dust, since in that case $\rho'$ is a constant.

Also note that many
different clock Hamiltonians $m(\zeta,J)$
lead to equivalent results.
This is
because,
for any invertible function
$f(J)$, the replacement of $J$ by $\tilde J = f(J)$ and
$\Theta$ by $\tilde \Theta = \Theta /  f'(J)$ changes
the kinetic term (3.1) only by a boundary term
but changes
the clock Hamiltonian from $m(\zeta,J)$ to $m(\zeta,f(J))$.
Thus, any two clock Hamiltonians $m_1$ and $m_2$ related by
$m_1(\zeta,J) = m_2(\zeta,f(J))$ for invertible $f$ are
equivalent.

Finally, consider a homogeneous and isotropic reference system
with DeWitt--type clock coupling. As before, the internal energy
density has the form
$\u{w} = \sqrt{h} w(\u{n}/\sqrt{h})$,
and now the mass is a function of $J$ only: $m = m(J)$. {}From
Eq.~(\ref{SLE}), the action becomes
\begin{equation}
\label{SLE-HI}
S_{\sss LE}[Z^i,J,\Theta;\gamma_{\alpha\beta}] = \int_{\cal M} d^4y
\sqrt{-\gamma} \Bigl\{ \u{n}J\Theta,_\alpha U^\alpha /\sqrt{h} -
\rho(\u{n}/\sqrt{h},J) \Bigr\} \ ,
\end{equation}
where $\rho(n,J) = n m(J) + w(n)$. This action is equivalent to
the action (refered to as the `hybrid action' in Ref.~\cite{d2})
for a nonisentropic perfect fluid with equation of state
$\rho(n,J) = n m(J) + w(n)$. In the perfect fluid literature,
the variable $J$ is interpreted as the entropy per particle and
the variable $\Theta$ is interpreted as the thermasy \cite{Dantzig}.
(The thermasy is a variable whose gradient along the particle
worldlines is proportional to the local temperature.) The
connection between the action (\ref{SLE-HI}) above and the
action of Ref.~\cite{d2} can be established easily.
One simply uses Eqs.~(2.7) and (2.8) to show that the
quantity $\sqrt{-\gamma/h} \u{n} U^\alpha$ that appears in the
action (3.6) is the same function of $Z^i$ and
$\gamma_{\alpha\beta}$ as the quantity $J^\alpha$ that appears
in the `hybrid action' of Ref.~\cite{d2}.


\subsection{Hamiltonian formulation of reference systems}

We would like to study the diffeomorphism invariance of general
relativity by using the reference system defined by DeWitt's
method of coupling clocks to an elastic medium. Since we will
examine this issue from the canonical perspective, we first
perform a Hamiltonian analysis of the reference system. In order
to make contact with the usual
Hamiltonian description of gravity, we will work initially in the
Eulerian picture.

We begin by foliating spacetime ${\cal M}$ with
a family of hypersurfaces $\Sigma_t$ labeled by the parameter $t$.
If $x^a$ are coordinates on a hypersurface, then the spacetime
coordinates $y^\alpha$ are related to $(t,x^a)$ through mappings
$y^\alpha = Y^\alpha(t,x)$.
As usual, the lapse function $N^{\perp}$, shift vector $N^a$,
and spatial metric $g_{ab}$ are related to the spacetime metric by
\begin{eqnarray}
{\dot Y}^{\a} & = & N^{\perp}n^{\a}+ N^a Y^{\a},_a \ , \\
g_{ab} & = & Y^{\a},_a \gamma_{\a \b} Y^{\b},_b \ , \\
\gamma^{\a\b} & = & - n^\a n^\b + Y^{\a},_a g^{ab} Y^{\b},_b
\ ,
\end{eqnarray}
where $n^\a$ is the unit normal of the hypersurfaces. The quantities
$N^\perp$, $N^a$, $g_{ab}$, and $n^\a$ are functions of $t$ and $x^a$.
The  spacetime metric depends on $t$ and $x^a$ through the mapping
$Y^\alpha$; that is, $\gamma_{\a\b} =
\gamma_{\a\b}(Y(t,x))$. Note that we are now using the `dot' to
denote $\partial/\partial t$.

The variables $Z^i(y)$, $\Theta$, and $J$ are spacetime scalars. They
can be pulled back from ${\cal M}$ to $\Sigma\times\RR$ by the mapping
$Y^\alpha(t,x)$ to yield $t$--dependent scalars on
$\Sigma$ which we denote by the same kernel letter: for example,
$Z^i(y)\to Z^i(t,x)$. With the definitions above it is not difficult
to show that (with a slight abuse of notation)
\begin{eqnarray}
\label{Zids}
Z^i,_\a n^\a & = & ({\dot Z}^i - N^a Z^i,_a)/N^\perp \ , \\
Z^i,_\a Y^\a,_a & = & Z^i,_a \ , \\
Z^{i},{}^\a & = & -  n^\a Z^i,_\b n^\b + Y^\a,_a g^{ab} Z^i,_b \ .
\end{eqnarray}
Similar relations hold for $\Theta$ and $J$.

The function $Z^i(t,x)$ is a $t$--dependent mapping from the space
$\Sigma$ to the matter space $\cal S$.
It is useful to consider the inverse mappings $X^a(t,\zeta)$, defined
by $\zeta^i = Z^i(t,X(t,\zeta))$. The mappings $Z^i$ and $X^a$ are
related by the identities
$\delta^i_j = Z^i,_a X^a,_j$, $\delta^a_b = X^a,_i Z^i,_b$,
${\dot Z}^i = - Z^i,_a X^a,_t$, and ${\dot X}^a = - X^a,_i Z^i,_t$.
The notation used here is somewhat abbreviated
in that we have omitted any explicit
specification of the functional dependences. We will
continue this practice below.

Now consider the clock kinetic term (\ref{Ekin}).
{}From Eq.~(\ref{4v}) we have
\begin{equation}
\Theta,_\alpha U^\alpha = - {1 \over {3!}}  \epsilon_{\alpha
\beta \gamma \delta} \Theta,{}^\alpha Z^i,{}^\beta
Z^j,{}^\gamma Z^k,{}^\delta \epsilon_{ijk}
\end{equation}
which, using  Eq.~(3.12) and the identity
$n^\a \epsilon_{\alpha \beta \gamma \delta} Y^\b,_a
Y^\g,_b Y^\d,_c = \epsilon_{abc}$,
takes the form
\begin{eqnarray}
\Theta,_\a U^\a &=&  {1 \over {N^\perp}} (\dot{\Theta} - N^a \Theta,_a)
{1 \over {3!}} \epsilon^{abc} Z^i,_a Z^j,_b Z^k,_c \epsilon_{ijk} \cr
& &- {1 \over {2 N^\perp}} (\dot{Z}^i - N^a Z^i,_a)
\epsilon^{abc} \Theta,_a Z^j,_b Z^k,_c \epsilon_{ijk} \ .
\end{eqnarray}
It is convenient to write this result as
\begin{equation}
\Theta,_\a U^\a = \Gamma \dot{\Theta}/N^\perp + \Gamma(V^a -
N^a/N^\perp)\Theta,_a \ ,
\end{equation}
where
\begin{eqnarray}
\Gamma & \equiv & - n_\a U^\a =
{1 \over {3!}} \epsilon^{abc} Z^i,_a Z^j,_b Z^k,_c \epsilon_{ijk} \ ,\\
V^a & \equiv & {1\over N^\perp} ({\dot X}^a + N^a)
= -{1 \over {2 N^\perp\Gamma}}  \epsilon^{abc}
\epsilon_{ijk} Z^j,_b Z^k,_c ({\dot Z}^i - N^a Z^i,_a)  \ .
\end{eqnarray}
Here, $V^a$ is the spatial velocity of the particles as measured by the
observers who are at rest in $\Sigma_t$ and
$\Gamma = 1/\sqrt{1-V^a g_{ab} V^b}$ is the relativistic `gamma' factor
that characterizes the boost between these observers and the co--moving
reference frame.

Using
Eqs.~(2.7) and (3.16), we find that the fleet metric and its
determinant can be expressed as
\begin{eqnarray}
h^{ij} & = & Z^i,_a Z^j,_b (g^{ab} - V^a V^b) \ ,\\
\sqrt{h} & = & \sqrt{g} \Gamma / (\det Z^i,_a) \ .
\end{eqnarray}
These results, along with Eq.~(3.15), show that the action
(3.4) can be written as
\begin{eqnarray}
& & S[Z^i,J,\T;N^\perp,N^a,g_{ab}] \nonumber\\
& &\qquad = \int dt \int_{\Sigma} d^3x
(\det Z^i,_a) \biggl\{ \u{n} J {\dot \T}
+ \u{n}J(N^\perp V^a - N^a) \T,_a - N^\perp (\u{n} m + \u{w})/\Gamma
\biggr\} \ .
\end{eqnarray}
It is now straightforward to proceed with the canonical analysis.
The momentum conjugate to $\T$ is $(\det Z^i,_a)\u{n} J$, which we
will denote by $\Pi$. The momentum conjugate to $Z^i$ is
\begin{equation}
P_i \equiv \frac{\partial{\cal L}}{\partial {\dot Z}^i}  =
-(\det Z^j,_a) \Bigl[ \Gamma(\u{n}m+\u{w})V_b + \u{t}_{k\ell}
Z^k,_b Z^\ell,_c V^c/\Gamma + \u{n}J\T,_b \Bigr] X^b,_i
\ ,
\end{equation}
where $\u{t}_{ij} = 2 (\partial\u{w} / \partial h^{ij} )$ is
the Lagrangian frame stress tensor \cite{Bryce}.
Note that ${\dot Z}^i$
appears in this expression only in the combination
$V^a = -({\dot Z}^i X^a,_i - N^a)/N^\perp$. Thus, the solution
for ${\dot Z}^i$ as a function of the canonical variables, lapse,
and shift has the form
\begin{equation}
{\dot Z}^i = (N^a - N^\perp V^a) Z^i,_a \ ,\
\end{equation}
where we view $V^a$ as a function of
the canonical variables
through equations (3.21) and (3.18) for $P_i$ and $h^{ij}$.

Collecting together the above results, we find the Hamiltonian form of
the action for an elastic medium coupled to clocks in the
Eulerian picture:
\begin{equation}
S_{\sss HE}[Z^i,P_i,\T,\Pi;N^\perp,N^a,g_{ab}] = \int dt \int_\Sigma
d^3x \biggl\{ P_i {\dot Z}^i + \Pi {\dot \T} - N^\perp
{\cal H}^{\rm m}_\perp - N^a {\cal H}_a^{\rm m} \biggr\}
\ ,
\end{equation}
where
\begin{eqnarray}
{\cal H}_a^{\rm m} & = & P_i Z^i,_a + \Pi \T,_a \ ,\\
{\cal H}^{\rm m}_\perp & = & (\det Z^i,_a)
[ \Gamma(\u{n}m + \u{w}) + \u{t}_{ij}
Z^i,_a V^a Z^j,_b V^b /\Gamma ]  \ .
\end{eqnarray}
Here, $\Gamma$ is defined in Eq.~(3.16) and the matter
spatial velocity $V^a$ and fleet metric $h_{ij}$ are defined as
functions of the canonical variables through Eqs.~(3.18) and (3.21).
The clock Hamiltonian, number density, and interaction energy density
have dependences $m=m(\Pi/\u{n}(\det Z^i,_a),Z^j)$,
$\u{n} = \u{n}(Z^i)$, and $\u{w} = \u{w}(h_{ij},Z^k)$.


\section{Reference Systems in General Relativity}

Thus far we have treated the gravitational field as an external
background.  In this section, we shall include gravity as a dynamical
field and study our reference system
as coupled to general relativity.

\subsection{Coupling clocks to the gravitational field}

Since the action for the matter/clock system contains no
derivatives of the spacetime metric, the action for that system
coupled to gravity is obtained by adding its action to the
gravitational action. In the Eulerian setting, we have
\begin{eqnarray}
& &S_{\sss HE}[Z^i,P_i,\T,\Pi,g_{ab},p^{ab},N^\perp,N^a] \nonumber\\
& &\qquad\qquad = \int dt
\int_\Sigma d^3x \biggl\{ P_i {\dot Z}^i + \Pi {\dot \T}
+ p^{ab}{\dot g}_{ab}
- N^\perp ({\cal H}^{\rm m}_\perp + {\cal H}^{\rm g}_\perp)
- N^a ({\cal H}_a^{\rm m} + {\cal H}_a^{\rm g})
\biggr\} \ ,
\end{eqnarray}
where ${\cal H}^{\rm g}_\perp$ and ${\cal H}_a^{\rm g}$ are the
familiar constraints of vacuum general relativity \cite{ADM}.
The coupled system is subject to the constraints
\begin{eqnarray}
{\cal H}_\perp \equiv {\cal H}_\perp^{\rm m}
+ {\cal H}_\perp^{\rm g} = 0 \ ,\\
{\cal H}_a \equiv {\cal H}_a^{\rm m}
+ {\cal H}_a^{\rm g} = 0 \ ,
\end{eqnarray}
since the lapse function $N^\perp$ and shift vector $N^a$ are
varied in the action principle.

It is also useful to consider the coupled system in the
Lagrangian picture. The Hamiltonian form of the action,
namely $S_{\sss HL}$, can be obtained from a 3+1 decomposition of
the Lagrangian form of the action in the Lagrangian picture,
namely $S_{\sss LL}$. Alternatively, $S_{\sss HL}$ can be obtained
{}from the Eulerian picture action $S_{\sss HE}$ by performing a point
canonical transformation from Eulerian to Lagrangian variables.
The details of this transformation are described in Ref.~\cite{d2} for
the perfect fluid case and are essentially the same here.
Thus, we define the new matter variables $X^a(\zeta)$ such that
$Z^i(X(\zeta)) = \zeta^i$. We also define the new clock variable
$\Theta(\zeta) = \Theta(X(\zeta))$
and the new gravitational field variable
\begin{equation}
g_{ij}(\zeta) = X^a,_i(\z) X^b,_j(\z) g_{ab}(X(\z)) \ .
\end{equation}
This is the spatial metric expressed in the Lagrangian system of
coordinates induced on the slices $\Sigma$ by the matter.
The new conjugate momenta
are defined
by \cite{d2}:
\begin{eqnarray}
\u{P}_a(\z) & = & -(\det X^b,_j) {\cal H}_a(X(\z)) \ ,\\
\u{\Pi}(\z) & = & (\det X^b,_j) \Pi(X(\z)) \ ,\\
\u{p}^{ij}(\z) & = & (\det X^c,_k) Z^i,_a(X(\z)) Z^j,_b(X(\z))
p^{ab}(X(\z)) \ .
\end{eqnarray}
With this transformation, the action (4.1) becomes
\begin{eqnarray}
& & S_{\sss HL}[X^a,\u{P}_a,\T,\u{\Pi},g_{ij},\u{p}^{ij},N^\perp,N^a]
\nonumber\\ & & \qquad\qquad\qquad
= \int dt \int_{\cal S} d^3\z \biggl\{ \u{P}_a {\dot X}^a
+ \u{\Pi}{\dot\T} + \u{p}^{ij}{\dot g}_{ij}
- N^\perp(X) \u{{\cal H}}_\perp + N^a(X)
\u{P}_a \biggr\} \ ,\
\end{eqnarray}
where the lapse and shift are functions of $X^a(\z)$ and the
Hamiltonian constraint is $\u{{\cal H}}_\perp
= (\det X^a,_i){\cal H}_\perp$. In terms of the new canonical
variables, we have
\begin{equation}
\u{{\cal H}}_\perp = \Gamma(\u{n}m + \u{w}) + \u{t}_{ij}
Z^i,_a V^a Z^j,_b V^b /\Gamma + \u{{\cal H}}^{\rm g}_\perp
\ .\
\end{equation}
Here, we have used $\u{\cal H}^{\rm g}_\perp(\z) =
(\det X^a,_i){\cal H}^{\rm g}_\perp(X(\z))$ to denote the pullback
[by the mapping $X(\z)$] of
the gravitational contribution to the Hamiltonian constraint from
$\Sigma$ to ${\cal S}$. Since the
Lagrangian gravitational variables $g_{ij}(\z)$ and $\u{p}^{ij}(\z)$
are obtained from the Eulerian gravitational variables $g_{ab}(x)$
and $p^{ab}(x)$ by the pullback mapping from $\Sigma$ to ${\cal S}$,
the term $\u{\cal H}^{\rm g}_\perp$ is constructed by replacing the
Eulerian variables in ${\cal H}^{\rm g}_\perp$ with the corresponding
Lagrangian variables. Likewise, we can define the pullback of the
gravitational contribution to the momentum constraint by
$\u{\cal H}_i^{\rm g}(\z) =
(\det X^b,_j)X^a,_i{\cal H}_a^{\rm g}(X(\z))$.

The velocity $V^a$ and fleet metric $h_{ij}$ that appear in
the Hamiltonian constraint (4.9) must, of course, be expressed in terms
of the new canonical variables. It turns out to be convenient to
work with $V_a X^a,_i$, which is the velocity expressed in the Lagrangian
coordinate system, in place of $V^a$. In Appendix \ref{apptwo} we
show that $h_{ij}$ and $V_a X^a,_i$ can be expressed in
terms of $\u{\Pi}$, $\u{P}_a X^a,_i$, $g_{ij}$, and $\u{p}^{ij}$.

In the action (4.8), the lapse function and shift vector are
functions of $X^a(\z)$. We can define the new variables
$N^\perp(\z)$ and $N^i(\z) = Z^i,_a N^a(X(\z))$ by mapping
$N^\perp(x)$ and $N^a(x)$ from $\Sigma$ to $\cal S$.  The
action then reads
\begin{eqnarray}
& & S_{\sss HL}[X^a,\u{P}_a,\T,\u{\Pi},g_{ij},\u{p}^{ij},N^\perp,N^i]
\nonumber\\ & & \qquad\qquad\qquad\quad
= \int dt \int_{\cal S} d^3\z \biggl\{ \u{P}_a {\dot X}^a
+ \u{\Pi}{\dot\T} + \u{p}^{ij}{\dot g}_{ij}
- N^\perp \u{{\cal H}}_\perp - N^i
\u{\cal H}_i \biggr\} \ ,\
\end{eqnarray}
where the Hamiltonian and momentum constraints are
\begin{eqnarray}
\u{{\cal H}}_\perp & = & \Gamma(\u{n}m + \u{w}) + \Gamma^3 \u{t}^{ij}
V_a X^a,_i V_b X^b,_j + \u{{\cal H}}^{\rm g}_\perp \ ,\\
\u{\cal H}_i & = & - \u{P}_a X^a,_i \ .\
\end{eqnarray}
Here, the Hamiltonian constraint $\u{\cal H}_\perp$ is obtained from
Eq.~(4.9) and the identity (B2) of Appendix \ref{apptwo}. The momentum
constraint $\u{\cal H}_i$
equals the result obtained by mapping ${\cal H}_a$ to the matter
space ${\cal S}$; see Eq.~(4.5).

As a final remark, let us point out that there are two symmetries that
play an important role in the description of reference systems
\cite{BK,d1,d2}.
The first is the freedom to reset the clocks. That is,
the action is invariant under a transformation in which the zero
point of each clock is changed. The infinitesimal version of this
transformation is generated through the Poisson brackets by the functional
\begin{equation}
Q[\vartheta] = \int_{\cal S} d^3\z \, \vartheta \u{\Pi}
= \int_{\Sigma} d^3x \, \vartheta(Z(x)) {\Pi}(x)
\ .\
\end{equation}
Thus, the clock at $\z^i$ is transformed according to
$\delta\T(\z) \equiv \{ \T(\z),Q[\vartheta] \} = \vartheta(\z)$.
The canonical variables in the Lagrangian picture,
other than $\T(\z)$, remain unchanged under this transformation.
In the Eulerian picture, we have $\delta\T(x) = \vartheta(Z(x))$ and
$\delta P_i(x) = -\vartheta,_i(Z(x)) \,\Pi(x)$, with the other
canonical  variables unchanged.
The second symmetry is an invariance with respect to changes
in the Lagrangian coordinate labels. That is, the action is
invariant under diffeomorphisms of the matter space ${\cal S}$.
The infinitesimal version of this transformation is generated by
\begin{equation}
Q[\xi^i] = \int_{\cal S} d^3\z \, \xi^i
( \u{P}_a X^a,_i + \u{\Pi} \T,_i + \u{\cal H}^{\rm g}_i)
= - \int_\Sigma d^3x \, \xi^i(Z(x)) P_i(x) \ ,
\end{equation}
where the vector field $\xi^i(\z)$ is the infinitesimal generator of
the diffeomorphism.
The canonical variables in the Lagrangian picture just transform
according to their ${\cal S}$--tensor character. For the Eulerian
picture variables we find $\delta Z^i(x) = -\xi^i(Z(x))$ and
$\delta P_i(x) = \xi^j,_i(Z(x)) P_j(x)$, while the remaining variables
(which are tensors on $\Sigma$) are invariant.


\subsection{New constraints and the gravitational Hamiltonian}

When the Hamiltonian constraint can be resolved for the
clock momenta, a new set of constraints may be introduced that
allows the system to be deparametrized. We begin by recalling that,
as shown in Appendix B, the
fleet metric $h_{ij}$ and velocity $V_a X^a,_i$ are functions of
$\u{\Pi}$, $g_{ij}$, $\u{p}^{ij}$, and $\u{\cal H}_i
= -\u{P}_a X^a,_i$. It
follows that the Hamiltonian constraint (4.11) depends on the
gravitational canonical variables $g_{ij}$ and $\u{p}^{ij}$, the clock
momentum $\u{\Pi}$, and also the particle canonical variables in the
combinations given by the momentum constraints $\u{\cal H}_i$. We can set
$\u{\cal H}_i$ equal to zero in $\u{{\cal H}}_\perp$ without changing
the content of the constraints $\u{{\cal H}}_\perp = 0$,
$\u{\cal H}_i = 0$. Then the Hamiltonian constraint $\u{\cal H}_\perp$
depends only on $g_{ij}$, $\u{p}^{ij}$, and $\u{\Pi}$.

Let us assume that we can resolve the constraint $\u{{\cal H}}_\perp = 0$
with respect to $\u{\Pi}$.  We would then
obtain a new constraint that has the form
\begin{equation}
\u{{\cal H}}_\uparrow
= \u{\Pi} + \u{h}[g_{ij}, \u{p}^{ij}] \ ,
\end{equation}
where the true (gravitational) Hamiltonian $\u{h}$ is a functional of
the gravitational variables only. The constraints
$\u{{\cal H}}_\uparrow = 0$ and $\u{\cal H}_i = 0$ constitute
a complete set of constraints for the system which are
equivalent to the original Hamiltonian and momentum constraints.
We can smear $\u{{\cal H}}_\uparrow$ with a
prescribed function $N^\uparrow(\z)$ on $\cal S$ to form the functional
$H[N^\uparrow]$.
The smeared constraint $H[N^\uparrow]$ generates, through the Poisson
brackets, changes in the canonical variables that result when the
hypersurface $\Sigma$ is displaced along
the particle world lines by the clock time $N^\uparrow(\z)$.

The constraints $\u{{\cal H}}_\uparrow$ have vanishing Poisson
brackets among themselves and with the momentum constraints
$\u{\cal H}_i$. As usual, the momentum constraints form a
representation of the Lie algebra of spatial diffeomorphisms.
(Note that, since the gravitational
variables $g_{ij}$, $\u{p}^{ij}$ and clock variables $\T$, $\u{\Pi}$
are independent of $\Sigma$, they have vanishing Poisson brackets
with $\u{\cal H}_i$.) Since
there are no nonvanishing cross terms in the brackets
of $\u{{\cal H}}_\uparrow$ with itself, the gravitational
Hamiltonian $\u{h}$ must have vanishing brackets with itself:
$\{\u{h}(\z),\u{h}(\z') \}= 0$. We should emphasize that this
property holds for any gravitational Hamiltonian $\u{h}$ that can be
derived through the coupling to an elastic medium with clocks; that is,
for any choice of internal energy density $\u{w} = \u{w}(\zeta^i,h_{jk})$.
In practice, the equations that determine $\u{h}$ are not solvable by
analytical means except for the simplest of cases, such as dust
($\u{w} = 0$).\footnote{Some other matter couplings also generate
true Hamiltonians that depend only on the gravitational variables.
Examples that can be solved analytically include the massless
scalar field \cite{KR} and certain perfect fluids
with `bad' clock coupling \cite{d3}.}

Finally, note that the constraint (4.15) allows the system to be
deparametrized. That is, consider the action $S_{\sss HL}$ with
$N^\perp \u{\cal H}_\perp$ replaced by $N^\uparrow \u{\cal H}_\uparrow$.
The particle variables $X^a$ and $\u{P}_a$ are completely decoupled
{}from the gravitational and clock variables and can be dropped from the
action. The variables $N^\uparrow$ and $\u{\Pi}$ can be eliminated through
the solution of their equations of motion, and the parameter $t$ can be
replaced by the clock time $\theta = \Theta(t)$. This leads to the action
\begin{equation}
S[g_{ij},\u{p}^{ij}] = \int d\theta \int_{\cal S} d^3\zeta \Bigl\{
\u{p}^{ij} {\dot g}_{ij}  - \u{h}[g_{ij},\u{p}^{ij}] \Bigr\} \ ,
\end{equation}
where the dot denotes $\partial/\partial\theta$. This action for the
gravitational field contains no gauge (diffeomorphism) invariance.

\subsection{The case of dust}

We will now explicitly display the results for dust, where the
interaction energy density vanishes ($\u{w} = 0$).
In this case, as in Ref.~\cite{BK}, the
Hamiltonian constraint (4.11) becomes
\begin{equation}
\label{resC}
\u{{\cal H}}_\perp = \sqrt{ (\u{n} m)^2 + \u{\cal H}_i^{\rm g}
g^{ij} \u{\cal H}_j^{\rm g} } + \u{\cal H}_\perp^{\rm g}
\ .\
\end{equation}
Recall that $\u{n} = \u{n}(\z)$ and $m = m(\zeta,\u{\Pi}(\z)/\u{n}(\z))$.
Let us consider various possible choices for $\u{n}m$. With the clock
Hamiltonian $m(\zeta,J) = J/k$, where $k$ is a positive constant, then
$\u{n}m = \u{\Pi}/k$ and the new constraint takes the form
$\u{\cal H}_\uparrow = \u{\Pi} + \u{h}$ with the gravitational
Hamiltonian
\begin{equation}
\u{h} = \mp k\sqrt{ (\u{\cal H}_\perp^{\rm g} )^2 - \u{\cal H}_i^{\rm g}
g^{ij} \u{\cal H}_j^{\rm g} } \ .
\end{equation}
As described in Ref.~\cite{BK}, the presence of the square root
creates a serious difficulty for defining
a quantum theory. In particular,
it implies that the physical Hilbert space must be restricted to those
states for which the operator $(\hat{\u{\cal H}}_\perp^{\rm g} )^2
- \hat{\u{\cal H}}_i^{\rm g} \hat{g}^{ij} \hat{\u{\cal H}}_j^{\rm g}$ is
nonnegative, and the physical observables must be restricted to those
operators that keep the states in the physical Hilbert space. The
difficulty is that the obvious candidates for physical observables, the
metric $\hat{g}_{ij} = g_{ij}\times$
and its conjugate $\hat{\u{p}}^{ij}= -i\delta/\delta g_{ij}$ acting by
multiplication and differentiation, do not satisfy this criterion.

While different choices for the clock Hamiltonian are classically
equivalent, as long as they are defined for the same
range of $J$, they do not necessarily lead to the same quantum
theory.  Thus, we now investigate other possibilities.
With the clock Hamiltonian $m(\zeta,J) = \sqrt{J/k}$ (where we
assume $J\geq 0$), we find that
$\u{n}m = \sqrt{\u{n}\u{\Pi}/k}$ and the gravitational Hamiltonian
is
\begin{equation}
\u{h} = -k\Bigl[ (\u{\cal H}_\perp^{\rm g} )^2 -
\u{\cal H}_i^{\rm g} g^{ij} \u{\cal H}_j^{\rm g} \Bigr] / \u{n}
\ .
\end{equation}
This result suffers from the same difficulty as the
square root Hamiltonian (4.18) if we insist that the momentum $\u{\Pi}$
should be interpreted in terms of a real, nonnegative clock Hamiltonian
$m$. In that case $\u{\Pi}$ must be nonnegative (assuming the number
density $\u{n}$ is positive), and the new constraint
$\u{\cal H}_\uparrow = \u{\Pi} + \u{h} = 0$ implies that $\u{h}$
of Eq.~(4.19) must be nonpositive;
we again have the requirement
$(\hat{\u{\cal H}}_\perp^{\rm g} )^2 - \hat{\u{\cal H}}_i^{\rm g}
\hat{g}^{ij} \hat{\u{\cal H}}_j^{\rm g} \geq 0$ on the quantum theory.
On the other hand,
it is not obvious that the condition $\u{\Pi} \geq 0$ must be kept,
so one might argue that it should
be dropped. This leads to a quantum theory that is free from the
difficulties of the condition
$(\hat{\u{\cal H}}_\perp^{\rm g} )^2 - \hat{\u{\cal H}}_i^{\rm g}
\hat{g}^{ij} \hat{\u{\cal H}}_j^{\rm g} \geq 0$, but there
might arise a problem of interpretation for the theory since the
variables $\T$ and $\u{\Pi}$ would no longer have a simple interpretation
in terms of a real clock Hamiltonian.

Clearly, by choosing the clock Hamiltonian $m(\zeta, J)$ appropriately,
we can arrange for the gravitational Hamiltonian $\u{h}$ to be given by
any invertible function of $(\u{\cal H}_\perp^{\rm g} )^2 -
\u{\cal H}_i^{\rm g} g^{ij} \u{\cal H}_j^{\rm g}$ (with $\u{n}$ appearing
as necessary to keep the density weights balanced).   In particular,
for the clock Hamiltonian $m(\zeta, J) = (J/k)^{1/4}$ (again with
$J\geq 0$), we have $\u{n}m = {\u{n}}^{3/4} (\u{\Pi}/k)^{1/4}$ and
\begin{equation}
\u{h} = -k\Bigl[ (\u{\cal H}_\perp^{\rm g} )^2 -
\u{\cal H}_i^{\rm g} g^{ij} \u{\cal H}_j^{\rm g} \Bigr]^2 / (\u{n})^3
\ . \
\end{equation}
This choice for the gravitational Hamiltonian appears to avoid
the problems encountered in the other cases. In particular, the clock
momentum remains nonnegative, $\u{\Pi}\geq 0$, for all solutions of the
constraint $\u{\cal H}_\uparrow = 0$, even on
states for which $(\hat{\u{\cal H}}_\perp^{\rm g} )^2
- \hat{\u{\cal H}}_i^{\rm g} \hat{g}^{ij} \hat{\u{\cal H}}_j^{\rm g}$ is
a negative operator.
However, some difficulties with interpretation do arise if
we attempt to take seriously the constraint in its
original form $\u{\cal H}_\perp = 0$. In that case, for states satisfying
the constraints and on which
$(\hat{\u{\cal H}}_\perp^{\rm g} )^2
- \hat{\u{\cal H}}_i^{\rm g} \hat{g}^{ij} \hat{\u{\cal H}}_j^{\rm g}$ is
negative, we find that $m^2<0$.  Thus, for such states the clocks
are tachyonic.\footnote{Said differently, if the mass is still
defined by the real branch of $(J/k)^{1/4}$, it will not satisfy
$\u{\cal H}_\perp = 0$ for all solutions of the new constraint
${\u{\cal H}}_\uparrow = 0$.}
One might also be concerned that there is
something pathological about a Hamiltonian $\u{h}$ that
is eighth order in the gravitational momenta.



\acknowledgements
DM was supported by NSF grants PHY-9396246 and PHY-9008502
and by research funds
provided by the Pennsylvania State University.

\appendix

\section{Perfect fluids as fluids coupled to clocks}
\label{+clocks}

By making the substitution $\u{n} \to J\u{n}$ in the action (2.11)
and adding the kinetic term (3.2), we obtain an action for an
isentropic perfect fluid coupled to clocks:
\begin{equation}
\label{coup}
S[Z^i,J,\Theta;\gamma_{\alpha\beta}] = \int_{\cal M} d^4y
\Bigl[ -\sqrt{-\gamma} \rho({{J\u{n}}
/ {\sqrt{h}}}) + \u{n} J\Theta,_\a U^\a\sqrt{-\gamma/h}\Bigr] \ .
\ee
Here $h$ and $U^a$  are expressed in terms of the label fields
$Z^i$ through Eqs.~(\ref{ifm}) and (\ref{4v}), respectively. In this
appendix we show that this action is equivalent to the (isentropic)
`un--barred' perfect fluid action of Ref.~\cite{pf1,d1,d2}.
In notation consistent with the present work, that action is
\be
\label{unbar}
S[K^\a,\Theta, Z^i, \lambda_i;\gamma_{\a \b}] = \int_{\cal M} d^4y
\Bigl[ -\sqrt{-\gamma} \rho (|K|/\sqrt{-\gamma}) + K^\a (\Theta,_\a +
\lambda_i Z^i,_\a) \Bigr] \ ,
\ee
where $|K| \equiv \sqrt{-K^\alpha K_\alpha}$ is the norm of the
timelike, future directed, spacetime four--vector density $K^\a$.
(Here, our $K^\alpha$, $\lambda_i$, and $Z^i$ correspond respectively
to $J^a$, $\beta_{\sss A}$, and $\alpha^{\sss A}$ of Ref.~\cite{d1}
and to $J^\alpha$, $-W_k$, and $Z^k$ of Ref.~\cite{d2}.)

The essential step in comparing the actions (A1) and (A2) is to
note that the equations of motion for some of the fields in the action
(A.2) can be solved algebraically for those fields as unique
functions of the other fields.  These `dependent' fields
contain only redundant dynamical information and, when the
corresponding equations of motion are solved and the solutions
inserted into the action (\ref{unbar}), the resulting action is
equivalent to (\ref{unbar}).  We will see that this
procedure produces the action (\ref{coup}).

In order to identify the proper equations of motion to solve we
must first change coordinates on field space by writing
$K^\a = \kappa U^\a$, where $U^\a$ is a {\it unit} future pointing
timelike four--vector. Thus, the action (A2) can be written as
\be
\label{unbar2}
S[\kappa,U^\a,\Theta, Z^i, \lambda_i;\gamma_{\a \b}] = \int_{\cal M} d^4y
\Bigl[ -\sqrt{-\gamma} \rho (\kappa/\sqrt{-\gamma}) + \kappa U^\a (\Theta,_\a
+ \lambda_i Z^i,_\a) \Bigr] \ .
\ee
The variations in $U^\alpha$ must preserve
the normalization condition $U^\a U_\a =-1$.

Note in particular the equations of motion that follow from
varying $\lambda_i$ and $U^\a$:
\begin{eqnarray}
\label{vars}
{{\delta S} \over {\delta \lambda_i}} = 0 & \Rightarrow & U^\a Z^i,_\a
= 0 \ , \\
{\delta S \over \delta U^\alpha} = 0 & \Rightarrow &  \Theta,_\a
+ \lambda_i Z^i,_\a  = C U_\alpha \ .
\end{eqnarray}
Here, $C$ is a proportionality constant that arises due to the
restriction on $\delta U^\alpha$. These equations are easily solved
for $\lambda_i$ and $U^\alpha$. The solution for $U^\alpha$ is just
the expression (\ref{4v}). The solution for $\lambda_i$ is found by
contracting Eq.~(A5) with $\gamma^{\alpha\beta} Z^j,_\beta$ and
using Eq.~(A4). The result is
\begin{equation}
\label{lambda}
\lambda_i = - h_{ij} \Theta,_\alpha \gamma^{\alpha\beta} Z^j,_\beta
\ ,
\end{equation}
where the fleet metric is defined in terms of $Z^i$ by Eq.~(2.7).
The solution for $U^\alpha$ and $\lambda_i$ can be substituted into
the action (A3) to yield the equivalent action
\be
S[\kappa, \Theta, Z^i;\gamma_{\a\b}] = \int_{\cal M} d^4y \Bigl[
- \sqrt{-\gamma}
\rho(\kappa/\sqrt{-\gamma}) + \kappa U^\a \Theta,_\a \Bigr] \ ,
\ee
where $U^\a$ is the functional of $Z^i$ given by Eq.~(\ref{4v}).
If we now identify $\kappa = J \u{n} \sqrt{-\gamma/h}$ (another change
of coordinates on field space), this is just
the action (\ref{coup}).
We see that the `un-barred' fluid of Ref.~\cite{pf1,d1,d2} may be
interpreted as
an isentropic perfect fluid
coupled to a fleet of clocks.

\section{Fleet metric and matter velocity as functions of the
canonical variables}
\label{apptwo}

The velocity $V_a X^a,_i$ and fleet metric $h_{ij}$ that appear in
the Hamiltonian constraint (4.11) are expressed in terms of the
Lagrangian picture canonical variables as follows. First, combine
Eq.~(3.21) for $P_i$
with Eq.~(4.5) for $\u{P}_a$ to obtain
\begin{equation}
\u{P}_a = \Gamma (\u{n}m + \u{w}) V_a + \u{t}_{ij}
Z^i,_a Z^j,_b V^b /\Gamma - (\det X^b,_j) {\cal H}_a^{\rm g} \ .
\eqnum{B1}
\end{equation}
By solving Eq.~(3.18) for $g^{ab}$ and inserting the result into
$Z^i,_a V^a = Z^i,_a g^{ab} V_b$, we obtain the identity
\begin{equation}
Z^i,_a V^a = \Gamma^2 h^{ij} V_a X^a,_j \ .\eqnum{B2}
\end{equation}
Then Eq.~(B1) becomes
\begin{equation}
\u{P}_a X^a,_i\, +\, \u{\cal H}_i^{\rm g} = \Gamma \Bigl[
(\u{n}m + \u{w}) \delta^j_i + \u{t}_{ik} h^{kj} \Bigr] V_a X^a,_j
\ ,\eqnum{B3}
\end{equation}
where $\u{\cal H}_i^{\rm g}$ is the pullback to ${\cal S}$ of the
gravitational contribution to the momentum constraint. With the
identity (B2), one can confirm that
\begin{equation}
h_{ij} = g_{ij} + \Gamma^2 (V_a X^a,_i) (V_b X^b,_j) \eqnum{B4}
\end{equation}
is indeed the fleet metric; that is, $h_{ij}$ is the inverse
of the inverse fleet metric (3.18).

Now observe that the gamma factor $\Gamma = 1/\sqrt{1-V^2}$
can be expressed in terms of $V_a X^a,_i$ and $h^{ij}$. This can
be seen by again solving Eq.~(3.10) for $g^{ab}$ and inserting
the result into $V^2 = V_a g^{ab} V_b$. This yields $V^2/\Gamma^2
= (V^a X^a,_i)h^{ij} (V_b X^b,_j)$, which can be solved for
$\Gamma$ as a function of $V_a X^a,_i$ and $h^{ij}$. We therefore
see that, in principle, Eqs.~(B3) and (B4) can be solved for $h_{ij}$
and $V^a X^a,_i$ as functions of the canonical variables. In particular,
$h_{ij}$ and $V_a X^a,_i$ depend on $\u{\Pi}$ [which is
contained in the argument of the clock Hamiltonian $m$ in Eq.~(B3)],
$\u{P}_a X^a,_i$ [which appears on the
left--hand side of Eq.~(B3)],  $g_{ij}$ [which appears explicitly
in Eq.~(B4) and on the left--hand side of Eq.~(B3) in the
combination $\u{\cal H}_i^{\rm g}$] and $\u{p}^{ij}$ [which appears
on the left--hand side of Eq.~(B3) in the
combination $\u{\cal H}_i^{\rm g}$].


\end{document}